\documentclass[runningheads]{llncs}
\usepackage[T1]{fontenc}
\usepackage{graphicx}
\usepackage{hyperref}
\usepackage{pifont}
\usepackage{tabularx}
\usepackage{array, makecell}
\usepackage{cite}
\usepackage{amsmath,amssymb,amsfonts}
\usepackage{algorithm}
\usepackage{algorithmic}
\usepackage{textcomp}
\usepackage{xcolor}
\usepackage{url}
\usepackage{listings}
\usepackage{multirow}
\usepackage{booktabs}
\usepackage{subcaption}
\usepackage{caption}
\usepackage{colortbl}
\usepackage{marvosym}

\begin{document}
\title{LLM-Driven Online Aggregation for Unstructured Text Analytics}
%
%
\author{
Chao Hui\inst{1}\thanks{Equal contribution.} \and
Weizheng Lu\inst{2}\textsuperscript{*}\textsuperscript{(\Letter)} \and
Yanjie Gao\inst{2}\textsuperscript{*} \and
Lingfeng Xiong\inst{2} \and
Yunhai Wang\inst{2} \and
Yueguo Chen\inst{2}
}

\authorrunning{Hui et al.}

\institute{
Shandong University, Qingdao, China\\
\email{chaohui@mail.sdu.edu.cn}
\and
Renmin University of China, Beijing, China\\
\email{\{luweizheng,gaoyanjie,lenfeng2022,wang.yh,chenyueguo\}@ruc.edu.cn}
}

\newcommand{\toolname}{\text{OLLA}} 

\newcommand{\hc}[1]{{\textcolor{red}{\it [HC] #1}}}
\newcommand{\xlf}[1]{{\textcolor{violet}{\it [XLF] #1}}}
\newcommand{\lwz}[1]{\textcolor{blue}{\it [LWZ] #1}}
\newcommand{\yj}[1]{\textcolor{green}{\it [YJ] #1}}
\newcommand{\todelete}[1]{\textcolor{yellow}{\it [TODELETE] #1}}

\maketitle              
\begin{abstract}
Large Language Models (LLMs) exhibit strong capabilities in text processing, and recent research has augmented SQL and DataFrame with LLM-powered semantic operators for data analysis.
However, LLM-based data processing is hindered by slower token generation speeds compared to relational queries.
To enhance real-time responsiveness, we propose {\toolname}, an LLM-driven online aggregation framework that accelerates semantic processing
within relational queries. 
In contrast to batch-processing systems that yield results only after the entire dataset is processed, our approach incrementally transforms text into a structured data stream and applies online aggregation to provide progressive output.
To enhance our online aggregation process, we introduce a semantic stratified sampling approach that improves data selection and expedites convergence to the ground truth.
Evaluations show that {\toolname} reaches 1\% accuracy error bound compared with labeled ground truth using less than 4\% of the full-data time. It achieves speedups ranging from 1.6$\times$ to 38$\times$ across diverse domains, measured by comparing the time to reach a 5\% error bound with that of full-data time.
We release our code at \url{https://github.com/olla-project/llm-online-agg.git}.

\keywords{Large Language Model  \and Online Aggregation \and Text Processing.}
\end{abstract}

\section{Introduction}
\vspace{-0.1in}

Extracting insights from the ever-growing volume of unstructured text has been a long-standing research problem\cite{chen2021Text,wankhade2022survey}.
Due to the strong semantic understanding capabilities, Large Language Models (LLMs) have recently emerged as a powerful paradigm for unstructured text analysis, and researchers are integrating LLM-based text analysis into relational queries, with recent examples including LOTUS~\cite{patel2025Semantic} and UQE~\cite{dai2024UQE}.
These systems employ LLMs to parse unstructured text into structured fields and then perform relational analysis on the resulting data.
This approach enables users to perform complex statistical analysis on text using familiar interfaces such as SQL or DataFrame. 
However, converting text into structured fields requires row-by-row processing, and LLM token generation is significantly slower than the execution speed of relational queries~\cite{liu2025Optimizing}. 
For instance, processing a set of texts with an LLM might take several minutes, whereas a relational query may complete in seconds. 
This performance disparity hinders the large-scale adoption of LLM-based text analysis in production environments.
Thus, simply integrating LLMs into relational engines is insufficient to meet the demand for low-latency text analysis.

Inspired by online aggregation in relational analysis, which provides progressively refined approximate results, we propose a novel approach that combines online aggregation with LLM-based text analysis. 
We present \textbf{\toolname} (\textbf{O}nline \textbf{L}arge \textbf{L}anguage model \textbf{A}ggregator), a novel LLM-driven online aggregation framework to support fast, interactive analytics over large-scale unstructured text.
{\toolname} applies LLM-driven information transformation to convert unstructured data into a structured data stream and then performs online aggregation on the data stream.
To ensure our online aggregation process converges rapidly to the ground truth,
{\toolname} introduces a semantic stratified sampling strategy. 
It first converts unstructured text into a vector space using an embedding model, then clusters the embedding vectors, and finally samples and adjusts the resulting clusters.

Our evaluation demonstrates that {\toolname} reaches the 1\% absolute error bound on accuracy against the labeled ground truth using less than 4\% of the full data processing time. 
And it delivers speedups from 1.6$\times$ to 38$\times$, where speedups are measured by comparing the time required to reach a 5\% error bound of confidence interval with that of full data processing.

In summary, our contributions
are as follows:

\begin{itemize}

  \item {\toolname} builds upon the principles of online aggregation, enabling early, progressively refined query results. This allows users to obtain approximate insights in real time without waiting for full data processing to complete.

  \item We introduce a semantic indexing and stratified sampling mechanism where unstructured texts are converted into embedding vectors, clustered into strata, and then sampled uniformly from each stratum. This dynamic process improves both precision and efficiency over time.

  \item We implement a prototype and evaluate it on a range of representative queries over diverse real-world unstructured datasets. Experimental results show that our approach significantly outperforms baseline methods in terms of both response latency and convergence speed.
\end{itemize}

\vspace{-0.1in}

\section{Background \& Related Work}
\label{sec:background}

\subsection{LLM for Data Processing}

Large Language Models (LLMs) have recently shown strong capabilities in extracting and understanding semantics from unstructured text. 
Some recent studies extend SQL~\cite{dai2024UQE} or DataFrame~\cite{patel2025Semantic} on unstructured data analysis by embedding semantic operators driven by LLMs. 
Likewise, production systems such as Google BigQuery\footnote{\url{https://cloud.google.com/blog/products/ai-machine-learning/llm-with-vertex-ai-only-using-sql-queries-in-bigquery}} and Databricks\footnote{\url{https://docs.databricks.com/aws/en/large-language-models/ai-functions}} have integrated LLM capabilities directly into their SQL APIs.
This type of query invokes the LLM on each row to process the text column. 
Users have to wait for long periods to get results when working with large datasets, and the latency limits practicality in large-scale settings. 
\vspace{-0.15in}

\subsection{Online Aggregation}

Online aggregation provides users with approximate results early, refining results progressively as more data is processed~\cite{hellerstein1997Online, li2018Online}. 
Unlike batch queries that delay output until full completion, online aggregation supports incremental computation and interactive exploration, and enables users to observe query convergence over time and terminate early once results achieve the expected accuracy, which significantly reduces the time-to-insight.
Systems like Ripple Joins~\cite{haas1999Ripple}, BlinkDB~\cite{agarwal2013BlinkDB}, and VerdictDB~\cite{park2018VerdictDB} have demonstrated the effectiveness in streaming and approximate query processing contexts.
\vspace{-0.13in}

\section{LLM-driven Online Aggregation}
\label{sec:system}





\subsection{Problem Definition}

Consider a table $T=\{x_i\}_{i=1}^{N}$ with mixed attributes: some columns contain unstructured text $\mathcal{D}_i$ (e.g., logs, reviews, conversations) while the rest are
numeric, categorical, or otherwise structured data.
A user query $Q$ combines ordinary SQL with an LLM-driven function $LLM(\mathcal{D}_i)$ and then applies an aggregate $g$ (e.g., \texttt{COUNT}, \texttt{AVG}).
Because invoking the LLM for every row is expensive, we incrementally aggregate, and at each step, provide a running estimate along with a $\alpha$-level confidence interval.

\subsection{Workflow of {\toolname}}

\begin{figure}[!ht]
    \centering
    \includegraphics[width=1.0\linewidth, keepaspectratio]{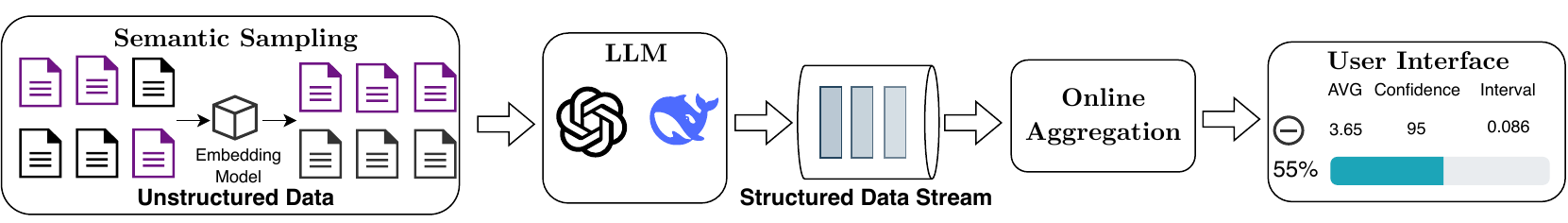}
    \caption{OLLA system architecture: unstructured data is processed by the LLM module to produce streaming structured data, which is then incrementally aggregated by the online aggregation engine.}
    \label{fig:overview}
    \vspace{-10px}
\end{figure}

To realize this vision of fast and interactive text analytics, {\toolname} integrates several key components, as illustrated in Fig.~\ref{fig:overview}.

\noindent \textbf{Semantic Stratified Sampling}.
In {\toolname}, each text entry is embedded into a high-dimensional vector space using an embedding model. 
The embedding vectors are then clustered using K-means algorithm~\cite{ahmed2020k}.
We then sample uniformly from the generated clusters. Since the initial embedding-based clusters may not align perfectly with the eventual LLM transformations, we introduce an adaptive adjustment algorithm. 
This algorithm iteratively adjusts and samples the strata, progressively purifying them to align with the LLM's output.

\noindent \textbf{Unstructured Data to Structured Data Streams}.
Raw unstructured data is transformed into a structured data stream using LLMs.





\noindent \textbf{Online Aggregation}.
{\toolname} adopts online aggregation to process 
incrementally arrived data streams, and the system delivers approximate answers with confidence bounds, enabling timely insights without waiting for full computation.


\subsection{Implementation}
\label{subsec:impl}

We built a demonstration system to verify the viability of our approach.
In the semantic sampling module, we adopt SentenceTransformers v4.1.0~\cite{reimers2019SentenceBert} and Faiss v1.8.0~\cite{johnson2019billion} to implement our sampling algorithm. 
We use the \textit{all-MiniLM-L6-v2} model~\footnote{\url{https://huggingface.co/sentence-transformers/all-MiniLM-L6-v2}} as our embedding model to encode text into dense vectors. 
In the unstructured-to-structured transformation module, we deploy an LLM inference service based on vLLM v0.8.4~\cite{kwon2023efficient}.
To ensure broad applicability, this module supports both on-premise (self-hosted) and cloud-hosted deployment configurations.
The output of the LLM service is continuously streamed to a Kafka v4.0.0~\cite{kreps2011kafka} message broker, which acts as a buffer and transport layer between the text transformation module and the downstream online aggregation engine. 
For the online aggregation and query processing, we use Apache Spark Streaming v3.5.1~\cite{armbrust2018Structured}.
Spark Streaming allows us to apply complex transformations and aggregations over real-time structured data, such as group-by or filtering operations. 
It also computes the confidence intervals.

\subsection{Query Examples}
\label{subsec:query_examples}

\begin{figure*}[!hbtp]
    \centering
    \includegraphics[width=0.98\linewidth]{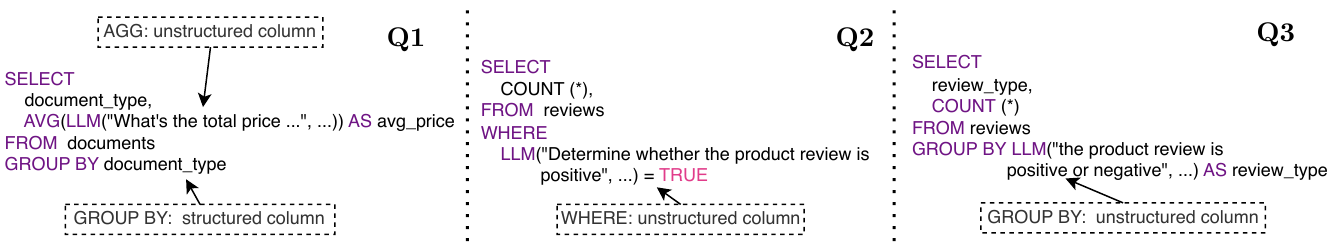}
    \caption{Three types of user queries.}
    \label{fig:query_type}
\vspace{-20px}
\end{figure*}

As shown in Figure~\ref{fig:query_type}, we categorize the user queries that can be accelerated by our method into three types:

Q1 - LLM in the \texttt{SELECT} clause: \texttt{GROUP BY} on structured columns with aggregation on an unstructured column using LLM. Traditional sampling is applied to the structured columns. We refer to this query type as \texttt{SELECT}.

Q2 - LLM in the \texttt{WHERE} clause: filters on an unstructured column and aggregation on structured columns. We call this query type \texttt{WHERE}.

Q3 - LLM in \texttt{GROUP BY} clause: \texttt{GROUP BY} an unstructured column and aggregates over structured columns. We denote this query type as \texttt{GROUP BY}.

We classify the three categories because online aggregation relies on data sampling, and the sample method directly affect the result's accuracy and the confidence interval's convergence speed~\cite{li2018Online}.
Sampling on structured columns can adopt conventional techniques, such as uniform sampling \cite{olken1995Random} or stratified sampling \cite{joshi2008Robust}.
Unstructured columns in \texttt{GROUP BY} or \texttt{WHERE} clauses cannot leverage traditional sampling techniques, and we will discuss them in the next Section~\ref{sec:semantic_sampling}

\section{Semantic Sampling \& Statistical Estimation}
\label{sec:semantic_sampling}



\subsection{Problem Formulation}
\label{subsec:formulation}

Given the text collection $\mathcal{D}$, we first construct a semantic embedding space through a mapping function $E: \mathcal{D} \rightarrow \mathbb{R}^d$, where each text is transformed into a $d$-dimensional dense vector.
The stratification process partitions the embedding space into strata using K-means clustering.
Each stratum $S_h$ is characterized by
$S_h = (\mathcal{D}_h, \mathcal{M}_h)$
where $\mathcal{D}_h \subset \mathcal{D}$ represents the contained texts and $\mathcal{M}_h$ maintains stratum-specific statistics.
For online estimation, we employ the following equation to get the confidence interval:
\begin{equation}
\label{eq:confidence_interval}
\epsilon_n = \left( \frac{z_p^2 V_n}{n} \right)^{1/2}
\end{equation}
where $z_p$ is the quantile value determined by the confidence level, $n$ is the sample size, and $V_n$ represents the variance term.
The specific computation of $p$ and $V_n$ varies between filtering and group aggregation scenarios.

The stratification achieves two primary objectives.
For \textit{filtering} cases, it helps identify and prioritize strata likely to contain valid samples (\texttt{WHERE TRUE}), thereby accelerating confidence interval convergence.
For \textit{group aggregation} cases, it uniformly samples from semantically coherent groups.
The estimator is unbiased in \textit{filtering} because it is derived from standard stratified random sampling with fixed strata. In \textit{group aggregation}, it is asymptotically unbiased because any potential bias from the dynamic stratum adjustment diminishes to zero as the sample size increases.

\subsection{Online Filtering with Semantic Stratified Sampling}
\label{subsec:filter_semantic_stratified}

In the \textit{filtering} scenario, our goal is to estimate aggregations over records that satisfy the \texttt{WHERE} clause. 
Two key observations motivate our sampling strategy. 
First, only valid samples, i.e., records for $\mathrm{LLM}(x) = \text{True}$, contribute to the final result, as data not satisfying the condition are discarded. 
Second, the confidence interval of the online aggregation narrows as more valid samples are collected, as described in Equation~\ref{eq:confidence_interval}. 
Therefore, by prioritizing early sampling from strata more likely to yield valid records, we can accelerate the convergence of the confidence interval.

Each stratum $S_h$ maintains a set of statistics
$\mathcal{M}_h = \{\hat{p}_h, \pi_h\}$
where $\hat{p}_h$ represents the estimated valid rate and $\pi_h$ indicates the sampling priority based on this rate. To compute these statistics, we first draw an initial total sample of size $m$ from the entire dataset $\mathcal{D}$, assuming it is representative of the overall data distribution. This sample is then allocated to each stratum $S_h$ proportionally to its size. The set of records sampled from stratum $S_h$ is denoted as $\mathcal{R}_h$, with a size of $m_h = |\mathcal{R}_h| = m \cdot |\mathcal{D}_h|/|\mathcal{D}|$. The valid rate is then estimated using this initial sample set:
\begin{equation}
\hat{p}_h = \frac{1}{m_h} \sum_{x \in \mathcal{R}_h} \mathbb{I}[\mathrm{LLM}(x) = \text{True}],
\end{equation}
where $\mathbb{I}[\cdot]$ denotes the indicator function. 
Strata are ranked in descending order of $\hat{p}_h$ to assign sampling priorities $\pi_h$, guiding the subsequent sampling process toward those more likely to yield valid records.
Let $v(x)$ denote the value of the aggregation column for a record with text $x$.
For example, if we aim to compute the average age of people whose textual descriptions meet certain criteria, then $v(x)$ returns the age value for the record containing text $x$.
The variance term $V_n$ in the confidence interval (Equation~\ref{eq:confidence_interval}) is computed as
\begin{equation}
V_n = \frac{1}{n-1} \sum_{i=1}^n (v(x_i) - \bar{v})^2,
\end{equation}
where $v(x_i)$ is the aggregation value of the $i$-th valid sample, and $\bar{v} = \frac{1}{n}\sum_{i=1}^n v(x_i)$ is their mean.
The confidence level $p=\alpha$ is typically pre-specified (e.g., 95\%).

\subsection{Online Aggregation with Semantic Stratified Sampling}
\label{subsec:semantic_groupby}

For \texttt{GROUP BY} on text, the main challenge is that group membership is unknown prior to LLM inference.
Our approach leverages the semantic similarity principle: texts with similar embeddings are likely to be classified into the same group. 
Based on this insight, we develop an adaptive stratification strategy that iteratively refines strata through sampling, recording, and adjustment phases.

Let $\mathcal{D}$ be a dataset with $N$ records, where each record belongs to one of $K$ possible categories after LLM inference. At the initial stage (i.e., $t=0$), the stratification of the dataset is generally performed based on prior knowledge or empirically. 
A common approach is to determine the number of strata $H$ using the total sample size $N$ and the number of categories $K$.
For instance, the empirical rule $H_0 =K\log N,H_{\text{max}}=2K \log N$ can be applied, where $H_0$ is the initial estimated number of strata and $H_{\text{max}}$ denotes a predefined maximum number of strata.
At time $t$, each stratum $S_h^{(t)}$ maintains a set of statistics:
\begin{equation}
\mathcal{M}_h^{(t)} = \{m_h^{(t)}, X_{hk}^{(t)}, \hat{p}_{hk}^{(t)}, \tau_h^{(t)}, \hat{V}_h^{(t)}\},
\end{equation}
where 
$m_h^{(t)}$ is the number of samples collected by time $t$, 
$X_{hk}^{(t)}$ tracks category $k$'s frequency, $\hat{p}_{hk}^{(t)} = X_{hk}^{(t)}/m_h^{(t)}$ estimates category $k$'s proportion, 
$\tau_h^{(t)} = \arg\max_{k} X_{hk}^{(t)}$ identifies the dominant category, 
and $\hat{V}_h^{(t)}$ measures stratum heterogeneity, defined as
\begin{equation}
\hat{V}_h^{(t)} = \sum_{k=1}^K \hat{p}_{hk}^{(t)}(1-\hat{p}_{hk}^{(t)}).
\end{equation}

\begin{figure}[!hbtp]
    \centering
    \includegraphics[width=1\linewidth]{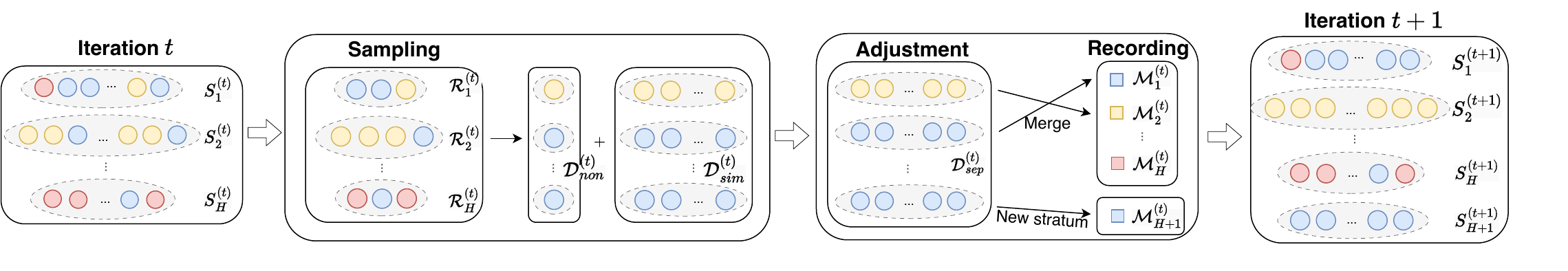}
    \caption{The adjustment process that enforces homogeneity within each stratum and heterogeneity across strata.}
    \label{fig:sampling_adjust}
\end{figure}

Our iterative process consists of three phases: Sampling, Recording, and Adjustment. Fig. \ref{fig:sampling_adjust} depicts these phases, which we discuss in detail below.

\textbf{Sampling}.
In each iteration $t$, we draw a total of $n^{(t)}$ samples across all strata, denoted as $\{\mathcal{R}_h^{(t)}\}_{h=1}^H$. For each stratum $S_h^{(t)}$, the sample size $n_h^{(t)} = |\mathcal{R}_h^{(t)}|$ is determined by Neyman allocation:
\begin{equation}
n_h^{(t)} = 
\begin{cases}
   n^{(t)} \cdot \frac{N_h^{(t)}}{N}, & \text{if } t = 0, \\
   n^{(t)} \cdot \frac{N_h^{(t)} \sqrt{\hat{V}_h^{(t)}}}{\sum_{i=1}^H N_i^{(t)} \sqrt{\hat{V}_i^{(t)}}}, & \text{if } t > 0.
\end{cases}
\end{equation}
The Neyman allocation method minimizes the variance of stratified sampling theoretically by considering both the stratum size and its variance, 
ensuring that strata with higher variability are allocated more samples to improve the precision and efficiency of the estimates. \cite{neyman1992Two}
At the initial iteration ($t = 0$), the sample size is allocated proportionally to the stratum size. 
In subsequent iterations ($t > 0$), the allocation takes into account both the stratum size and its estimated variance $\hat{V}_h^{(t)}$.

\textbf{Recording}.
After obtaining samples $\{\mathcal{R}_h^{(t)}\}_{h=1}^H$, we stream them to LLM for inference. 
As responses arrive, we update the stratum statistics $\mathcal{M}_h^{(t)}$ by recording frequency counts $X_{hk}^{(t)}$, computing proportions $\hat{p}_{hk}^{(t)}$, identifying dominant category $\tau_h^{(t)}$, and measuring heterogeneity $\hat{V}_h^{(t)}$.

\textbf{Adjustment}.
Each stratum $S_h^{(t)}$'s normalized variance is
\begin{equation}
\tilde{V}_h^{(t)} = \frac{\hat{V}_h^{(t)}}{1 - 1/K}.
\end{equation}
Here, the denominator $1 - 1/K$ corresponds to the theoretical maximum variance under a uniform category distribution, serving as a normalization factor.
If the normalized variance exceeds threshold $\theta$, we trigger a refinement process to improve stratum homogeneity by first identifying heterogeneous groups:
\begin{equation}
\mathcal{D}_{non}^{(t)} = \{x \in \mathcal{R}_h^{(t)} \mid \text{label}(x) \neq \tau_h^{(t)}\},
\end{equation}
and then locating semantically similar, unsampled records:
\begin{equation}
\mathcal{D}_{sim}^{(t)} = \{x \in \mathcal{D}_{h}^{(t)} \backslash \mathcal{R}_h^{(t)} \mid \exists y \in \mathcal{D}_{non}^{(t)}: \text{sim}(E(x),E(y)) > \gamma\},
\end{equation}
where $\text{sim}(u,v)$ denotes the cosine similarity between embedding vectors $u$ and $v$.
These two sets are combined as $\mathcal{D}_{sep}^{(t)} = \mathcal{D}_{non}^{(t)} \cup \mathcal{D}_{sim}^{(t)}$, which represents potentially misclustered or outlier samples requiring reallocation.
Based on the number of existing strata, we either split or merge $\mathcal{D}_{sep}^{(t)}$.
When $H < H_{\text{max}}$, we create a new stratum from $\mathcal{D}_{sep}^{(t)}$.
Otherwise, we merge $\mathcal{D}_{sep}^{(t)}$ with an existing stratum sharing the same dominant category.
The target stratum $h^*$ is selected based on maximum embedding similarity:
\begin{equation}
h^* = \arg\max_{h'} \{\text{sim}(E(\mathcal{D}_{sep}^{(t)}), E(\mathcal{D}_{h'})) \mid \tau_{h'}^{(t)} = \tau_{\mathcal{D}_{sep}^{(t)}}\},
\end{equation}
where $E(\mathcal{D})$ represents the mean embedding vector of all texts in set $\mathcal{D}$.
After each iteration, we compute the variance term $V_n$ and confidence level $p$ for Equation~\ref{eq:confidence_interval}:
\begin{equation}
V_n = \sum_{h=1}^H \left(\frac{N_h^{(t)}}{N}\right)^2 f_h^{(t)} \hat{V}_h^{(t)},
\end{equation}
where $f_h^{(t)} = \frac{N_h^{(t)}-m_h^{(t)}}{m_h^{(t)}(N_h^{(t)}-1)}$ is the finite population correction term arising from the hypergeometric distribution due to sampling without replacement within each stratum, 
and $p = \alpha/K$ accounts for multiple comparisons across $K$ categories.

This iterative process continues until convergence (all strata achieve sufficient purity ($\tilde{V}_h^{(t)} \leq \theta$)), with each cycle improving stratum purity ($V^{(t+1)} < V^{(t)}$) while maintaining bounded variance ($0 \leq V_h \leq 1 - \frac{1}{K}$). 
The complete procedure is summarized in Algorithm~\ref{algo:sample_group}.

\begin{algorithm}[ht]
\caption{Online Aggregation with Semantic Stratified Sampling}
\label{algo:sample_group}
\begin{algorithmic}[1]
\REQUIRE Dataset $\mathcal{D}$, LLM function $\mathrm{LLM}$, Embedding function $E$, initial strata number $H_0$, maximum strata number $H_{\text{max}}$, confidence level $\alpha$, variance threshold $\theta$, similarity threshold $\gamma$

\FOR{each data point $x \in \mathcal{D}$}
    \STATE Compute embedding vector $e_x \leftarrow E(x)$
\ENDFOR
\STATE Partition embeddings $\{e_x\}$ into $H$ strata: $\{S_1^{(0)}, \dots, S_{H_0}^{(0)}\}$
\STATE $t \leftarrow 0$

\WHILE{$\exists h: \tilde{V}_h^{(t)} > \theta$}
    \IF{$t = 0$}
        \STATE $n_h^{(t)} \leftarrow n^{(t)} \cdot N_h^{(t)}/N$ for each stratum $S_h^{(t)}$
    \ELSE
        \STATE $n_h^{(t)} \leftarrow n^{(t)} \cdot N_h^{(t)}\sqrt{\hat{V}_h^{(t)}}/\sum_{i=1}^H N_i^{(t)}\sqrt{\hat{V}_i^{(t)}}$
    \ENDIF
    \STATE Draw samples $\mathcal{R}_h^{(t)}$ of size $n_h^{(t)}$ from each stratum
    
    \FOR{each sample $x \in \mathcal{R}_h^{(t)}$}
        \STATE Get category label $k \leftarrow \mathrm{LLM}(x)$
        \STATE Update $\mathcal{M}_h^{(t)} \leftarrow \{m_h^{(t)}, X_{hk}^{(t)}, \hat{p}_{hk}^{(t)}, \tau_h^{(t)}, \hat{V}_h^{(t)}\}$
    \ENDFOR
    
    \FOR{each stratum $h$ with $\tilde{V}_h^{(t)} > \theta$}
        \STATE Obtain $\mathcal{D}_{sep}^{(t)} \leftarrow \mathcal{D}_{non}^{(t)} \cup \mathcal{D}_{sim}^{(t)}$
        \IF{$H < H_{\text{max}}$}
            \STATE Create new stratum from $\mathcal{D}_{sep}^{(t)}$
        \ELSE
            \STATE Merge $\mathcal{D}_{sep}^{(t)}$ with most similar existing stratum $S_{h^*}$
        \ENDIF
    \ENDFOR
    \STATE Update confidence intervals using $V_n$ and $p = \alpha/K$
    \STATE $t \leftarrow t + 1$
\ENDWHILE

\RETURN Proportion estimates $\{\hat{p}_{hk}^{(t)}\}$ and confidence intervals
\end{algorithmic}
\end{algorithm}
\vspace{-0.1in}

\section{Evaluation}
\label{sec:eval}

\subsection{Experimental Setup}
\label{subsec:experiment_setup}

\textbf{Datasets.}
We evaluate our system using a diverse set of datasets spanning product reviews, document understanding, and text classification. 
Due to the absence of standard benchmarks for LLM queries, we curated real-world datasets from diverse sources and formulated representative queries over them. 
Table~\ref{tab:datasets_overview} shows the dataset description, size, and the queries on them.

\setlength{\tabcolsep}{6pt} 
\renewcommand{\arraystretch}{1.1}

\begin{table}[t]
\centering
\caption{Datasets and queries used in the experiments. The \textit{Size} column ($n_{\text{rows}}$) indicates the number of rows used. The \textit{Query Type} column corresponds to the three categories illustrated in Fig.~\ref{fig:query_type} and Section~\ref{subsec:query_examples}, as the LLM operator in the \texttt{SELECT}, \texttt{FILTER}, or \texttt{GROUP BY} clause.}
\label{tab:datasets_overview}
\scriptsize
\begin{tabular}{l m{0.5\textwidth} m{0.1\textwidth} l}
\toprule
\textbf{Dataset} & \textbf{Description} & \textbf{Size ($n_{\text{rows}}$)} & \textbf{Query Type} \\

\midrule
\makecell[l]{Company\\ Documents~\cite{companydocs2024}}& 
A document dataset comprising business documents like invoices, purchase orders.
& 10,000 & Q1 \\

\addlinespace 
\makecell[l]{BBC\\ News~\cite{learn-ai-bbc}} & 
A multi-class news classification dataset with synthetically added numerical fields (e.g., view counts) for aggregation.
& 15,000 & Q2,Q3 \\

\addlinespace

arXiv~\cite{arxiv-org-submitters-2024} & 
Contains metadata for research papers from arXiv, including titles and abstracts.
& 10,000 & Q2 \\

\addlinespace

\makecell[l]{Amazon\\ Product~\cite{he2016ups}} & 
Product reviews and metadata, with a sentiment distribution balanced for analysis. For our experiments, we created a balanced dataset with an equal number of positive and negative reviews.
& 15,000 & Q2 \\

\addlinespace

Movie~\cite{pang2005seeing} & 
A collection of movie reviews from critics, paired with metadata such as scores.
& 15,000 & Q2,Q3 \\

\addlinespace

\makecell[l]{Chinese\\ Resume~\cite{su2019resume}} & 
A collection of Chinese resumes where the self-introduction section includes personal information such as name and age.
& 20,000 & Q1 \\

\bottomrule
\end{tabular}
\vspace{-0.1in}
\end{table}

\noindent \textbf{Metrics.}
We adopt the statistical metrics: \textit{Absolute Error}, \textit{Confidence Interval}, and \textit{Cumulative Valids}. 
All reported times in our experiments refer to the average time per query.
The \textit{Absolute Error} between the progressive streaming output and the labeled ground truth can be referred to as accuracy.
The \textit{Confidence Interval} is computed following Equation~\ref{eq:confidence_interval},
with the specific computation of the confidence level $p$ and variance term $V_n$ varying between filtering and group aggregation scenarios as described in Section~\ref{sec:semantic_sampling}. 
The \textit{Cumulative Valids} refers to the total number of outputs that satisfy predefined validity constraints (e.g., syntactic correctness). 

\noindent \textbf{Environment.}
Our experiments are conducted on a machine equipped with 2 Intel Xeon 6438M CPUs (2.2 GHz, 60 MB cache), 512 GB of main memory, and 8 NVIDIA A800 GPUs (80 GB HBM memory), running Rocky Linux 8.9.

\vspace{-0.15in}

\subsection{Accuracy of {\toolname}}
\label{subsec:accuracy}

We evaluate the accuracy of {\toolname} along two dimensions: (1) the correctness of the LLM's unstructured-to-structured data transformation, and (2) the precision of the streaming aggregation result compared to the ground truth from the fully labeled dataset.

\begin{table}[hbtp]
\centering
\caption{Accuracy on \textit{BBC News} classification and \textit{Movie} sentiment analysis.}
\label{tab:model_accuracy_comparison_alt}
\scriptsize
\vspace{0.15in}
\begin{tabular}{lcc}
\toprule
\textbf{Model}      & \textbf{BBC News} & \textbf{Movie} \\
\midrule
Qwen2.5 7B          & 0.90              & 0.84           \\
DeepSeek-V3.1 671B       & 0.95              & 0.88           \\
Ground Truth        & 1.00              & 1.00           \\
\bottomrule
\end{tabular}
\vspace{-0.3in}
\end{table}

Table~\ref{tab:model_accuracy_comparison_alt} compares model accuracies on the \textit{BBC News} and \textit{Movie} datasets against their ground truth labels. 
The results indicate that LLMs performance is model-dependent and context-dependent, and LLMs' outputs do not 100\% align with the pre-defined labels.
The DeepSeek-V3.1 671B (no-thinking) model achieves the best performance, but it incurs a high inference cost. 
While the much smaller compact Qwen2.5-7B model is a compelling and cost-effective alternative, achieving accuracy rates of over 80\% and even 90\%. 

\begin{figure}[hbtp]
    \centering
    \includegraphics[width=0.7\linewidth]{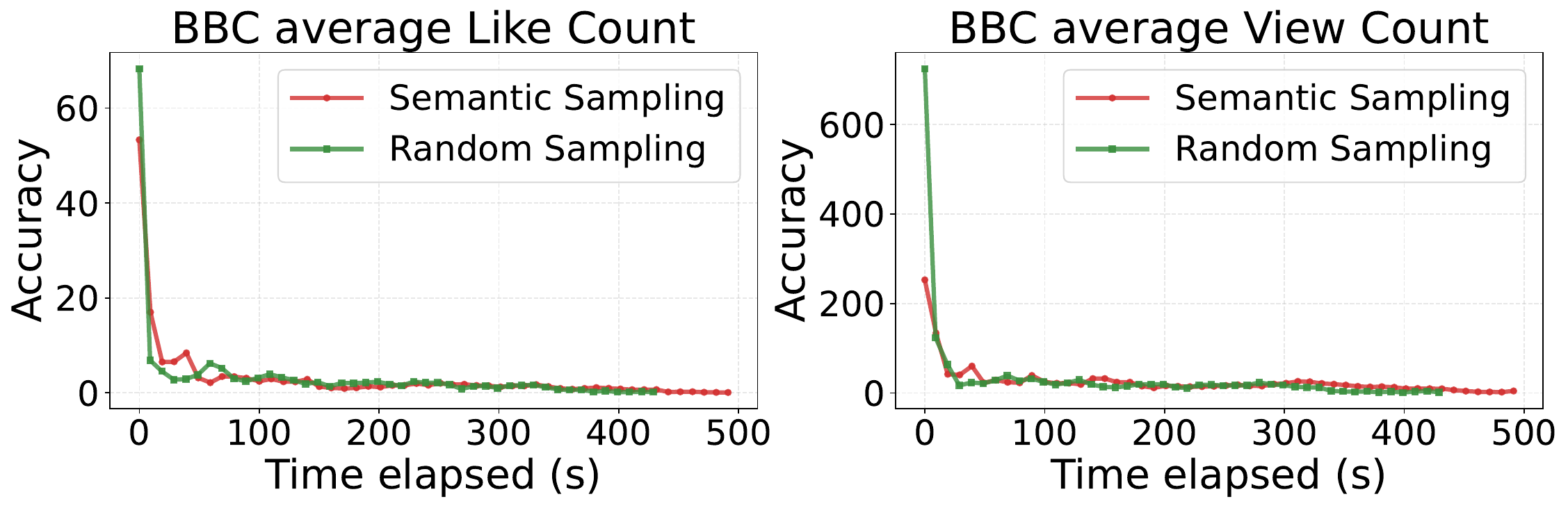}
    \vspace{-0.1in}
    \caption{Convergence of accuracy over time for online aggregation. Accuracy: the absolute error between the streaming aggregate result and the ground truth. Left: average \textit{like\_count} of the \textit{BBC News}. Right: average \textit{view\_count} of the \textit{BBC News}.}
    \label{fig:exp_acc}
\vspace{-0.3in}
\end{figure}

We compare the online aggregation results against the ground truth from the full dataset to directly measure {\toolname}'s empirical performance, rather than relying solely on its confidence interval estimates.
Fig.~\ref{fig:exp_acc} plots the absolute error between the streaming aggregated result and the ground truth computed over the entire dataset, showing that the online aggregation process converges remarkably quickly. {\toolname} reaches the 1\% error bound comparing the final labeled ground truth using only 2.43\% and 3.49\% of the entire dataset processing time in our two settings.
This figure illustrates the fundamental trade-off in online aggregation: as execution time (x-axis) increases, the error decreases. 
What's more, our semantic sampling method reduces error more quickly than random sampling, particularly during the initial stages of the aggregation process.

\subsection{Efficiency of {\toolname}}
\label{subsec:efficiency_olla}

\begin{figure*}[ht]
    \centering
    \includegraphics[width=1.0\linewidth]{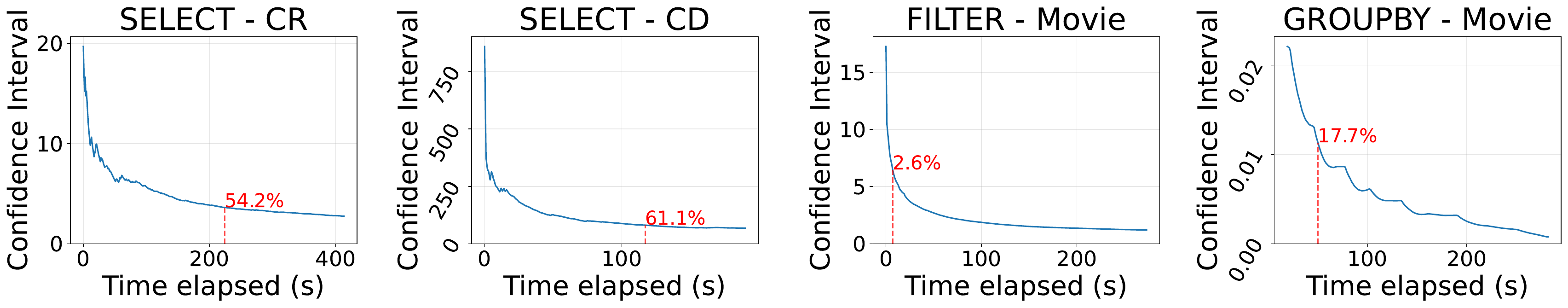}
    \caption{Convergence of confidence intervals for representative query types. The red percentage is the time our method reaches a 5\% error bound, divided by the total batch execution time. From left to right: \texttt{SELECT}: Average \textit{age} of resume on \textit{Chinese Resume} (CR); \texttt{SELECT}: Average \textit{total\_price} of type \textit{Invoices} on \textit{Company Documents} (CD); \texttt{WHERE}: Average length of positive \textit{Movie} reviews; \texttt{GROUP BY}: Proportion of Neutral \textit{Movie} reviews.}
    \label{fig:exp_base}
    \vspace{-0.1in}
\end{figure*}

\begin{figure}[ht]
    \centering
    \includegraphics[width=0.75\linewidth]{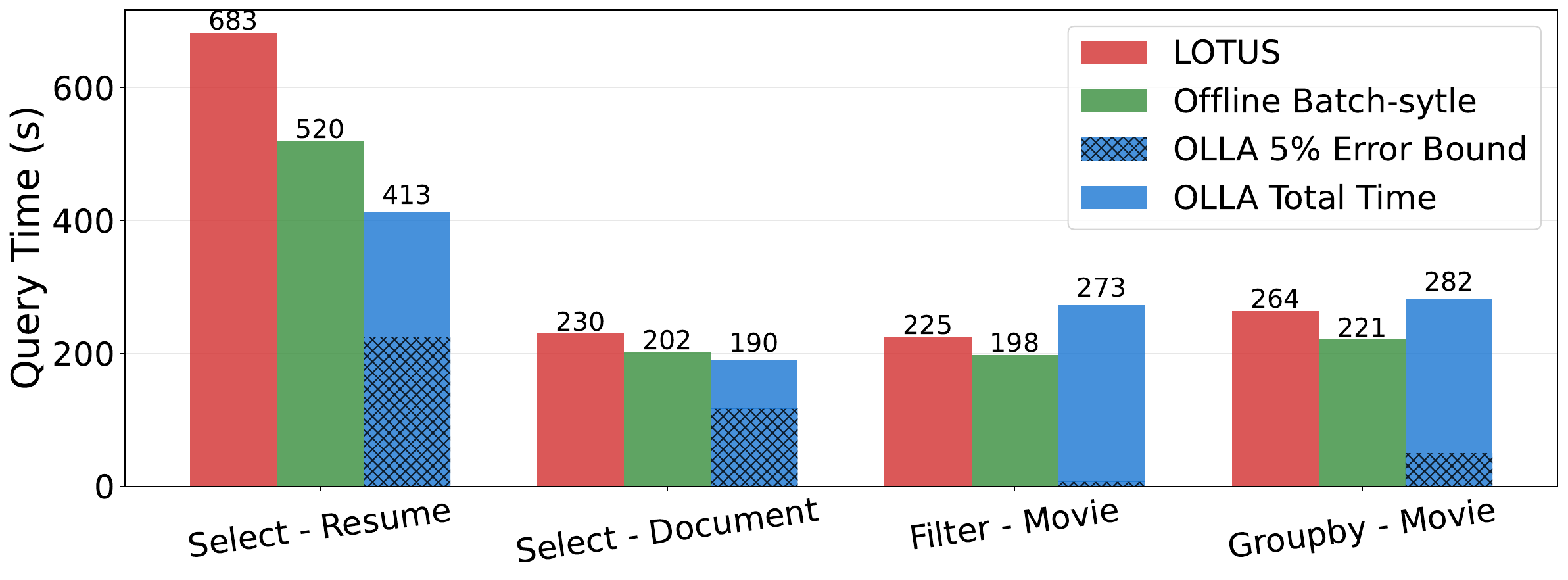}
    \vspace{-10pt}
    \caption{Comparison of query execution times for LOTUS, a batch-style baseline, and OLLA are shown for different query types. The hatched portion of the OLLA bar indicates the time to reach a 5\% error bound, while the full bar represents its total execution time.}
    \label{fig:exp_base_time}
\vspace{-0.2in}
\end{figure}

\vspace{-0.2in}

To validate the efficiency of {\toolname}, we conduct experiments across three representative query types illustrated in Fig.~\ref{fig:query_type}. 
In each case, we track the evolution of the confidence interval as the sampling progresses. 
This setup reflects a common use case where users seek to get early, progressively refined results without waiting for the full dataset to be processed. 


Fig.~\ref{fig:exp_base} presents the confidence interval trajectories for each query. We further evaluate the fraction of total execution time needed to reach a 5\% error bound. It is approximately 61.1\% (1.6$\times$ compared with scanning full data) of the total execution time for the \texttt{SELECT} scenario, 2.6\% (38$\times$) for the \texttt{WHERE} scenario, and 17.7\% (5.6$\times$) for the \texttt{GROUP BY} scenario.

We further conduct a comparative analysis of {\toolname} against both a state-of-the-art system, LOTUS~\cite{patel2025Semantic}, and a naive offline baseline that processes the entire data via parallel random sampling. 
The results, presented in Fig.~\ref{fig:exp_base_time}, highlight the trade-offs and core benefits of our approach. For the \texttt{SELECT} query, {\toolname} significantly reduces latency. For the \texttt{WHERE} and \texttt{GROUP BY} tasks, while our semantic stratified sampling introduces a minor latency overhead, the primary strength of {\toolname} lies in its online nature. The time required to reach the 5\% error bound is substantially shorter than waiting for any offline batch-style method to complete. 
\vspace{-0.15in}



\subsection{Efficiency of Sampling Optimization}

\textbf{Online Filtering with Semantic Sampling.}
\label{subsubsec:exp_filter_sampling}
We evaluate the effectiveness of online Filtering with Semantic Sampling on two datasets: \textit{Amazon Product} and \textit{BBC News}. In each experiment, we compare our method against a random sampling baseline in terms of how quickly they accumulate informative samples and how fast the confidence intervals converge during online aggregation.

\vspace{-0.2in}


\begin{figure}[ht]
    \centering
    \includegraphics[width=1.0\linewidth]{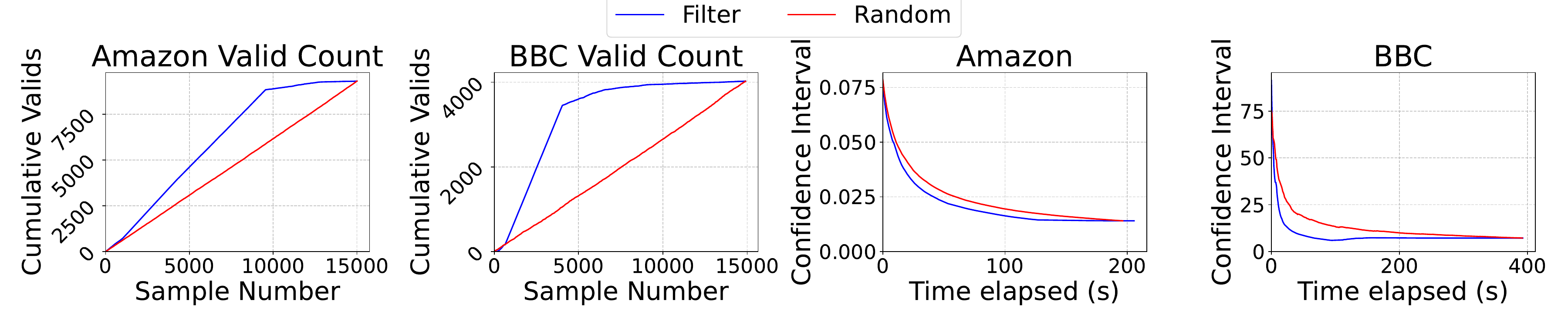}
    \vspace{-20pt}
    \caption{Comparison of Online Filtering with Semantic Sampling (``Filter" in figure) and random sampling (``Random" in figure). Left: \textit{verified\_purchase} rate among reviews classified as positive by the LLM in \textit{Amazon Product}; Right:  average \textit{like\_count} of news articles identified by the LLM as sports-related in \textit{BBC News}.}
    \label{fig:exp_filter}
\end{figure}

\vspace{-0.2in}

The results are presented in Fig.~\ref{fig:exp_filter}. The top two plots show the cumulative count of LLM-evaluated entries satisfying the predicate (i.e., labeled TRUE) as sampling steps. Our filter-aware strategy consistently accumulates valid samples more rapidly in the early phases of sampling.  
The bottom two plots show the width of the corresponding confidence intervals for the target aggregates, demonstrating that our method reaches the target error bound significantly faster than the baseline.
Specifically, for the \textit{Amazon} dataset, filter sampling reaches the 5\% error bound after retrieving 5.28\% of the full sampling process, while random sampling requires 6.44\%. Similarly, for \textit{BBC News}, filter sampling reaches this bound at 2.55\% of the process, in contrast to 6.04\% for random sampling, demonstrating that our method can provide tighter estimates with much lower sampling effort and, consequently, reduce the number of LLM calls in practice.

\noindent \textbf{Group Aggregation with Semantic Stratified Sampling.}
\label{subsubsec:exp_groupby_sampling}
We further assess the performance of Semantic Stratified Sampling in group-based aggregation scenarios, where the objective is to estimate the proportion or total of different groups as identified by the LLM. We conduct experiments on two classification tasks: categorizing \textit{Movie} entries by sentiment and \textit{BBC News} articles by topic.
Semantic Sampling proceeds in iterations, and the stratum is dynamically adjusted for every 10\% of the total sampling steps.
Our baselines include standard random sampling and the static stratification (i.e., do not adjust stratum during iterations) proposed in UQE~\cite{dai2024UQE}.


\begin{figure}[ht]
    \centering
    \includegraphics[width=1.0\linewidth]{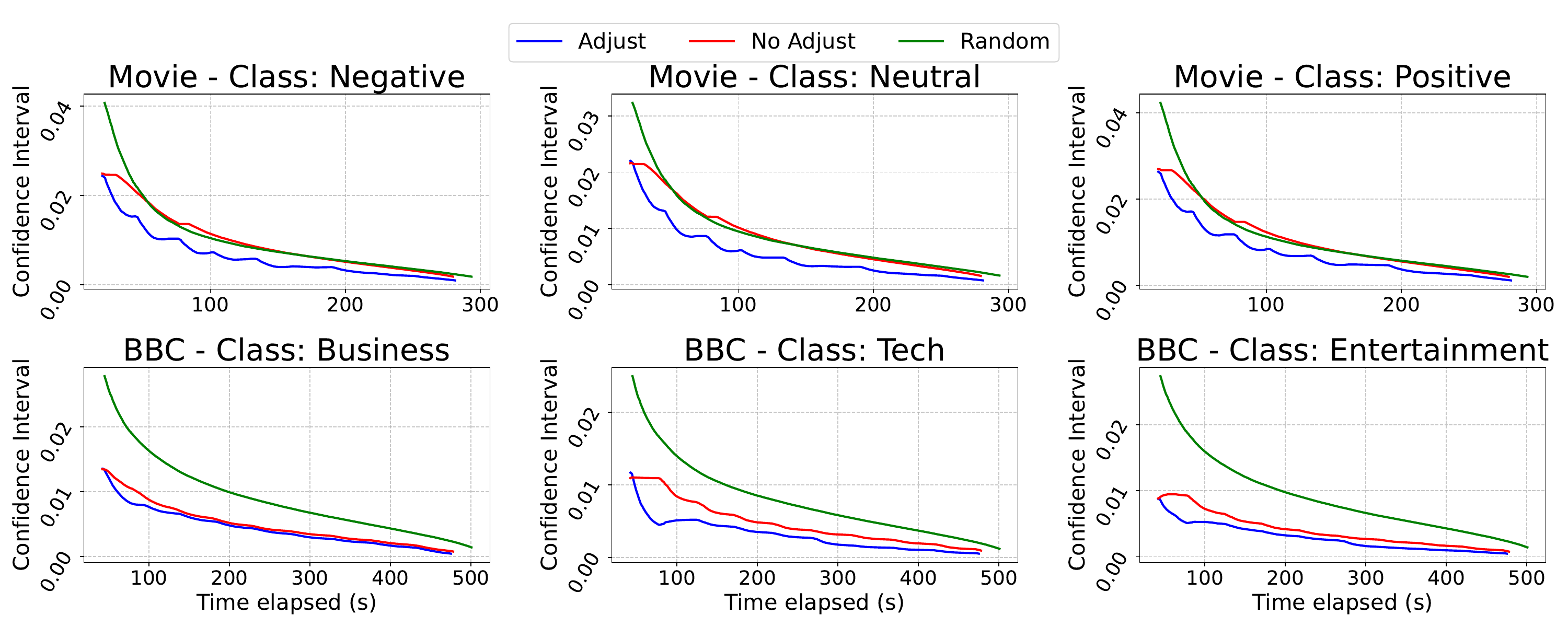}
    \vspace{-20pt}
    \caption{Confidence interval evolution for group proportion estimates under different sampling strategies. 
    Legend description: ``Adjust" – applies the Sampling-Recording-
    Adjustment workflow proposed in Section~\ref{subsec:semantic_groupby}; ``No Adjust" – performs sampling solely on the embedding vectors without adjusting the clustered strata proposed in UQE~\cite{dai2024UQE}; ``Random" – draws samples by simple random selection.
        Top: ratio of \textit{Movie} sentiment (Positive, Neutral, Negative); 
        Bottom: ratio of \textit{BBC News} topic (Business, Tech, Entertainment).}
    \label{fig:exp_dynamic}
\end{figure}

\vspace{-0.2in}

Fig.~\ref{fig:exp_dynamic} presents the results for both datasets. Across all cases, the \textit{Adjust} strategy, our dynamic stratified sampling method, consistently produces faster confidence interval convergence compared to the baselines. In the \textit{Movie} dataset, the static stratification (\textit{No Adjust}) performs similarly to random sampling, indicating that the initial clustering fails to align well with semantic labels. In contrast, the \textit{BBC News} dataset shows that even without adjustment, stratified sampling still outperforms random sampling, suggesting more effective initial cluster alignment. 
To quantitatively evaluate the efficiency of different sampling strategies, we report the average proportion of time required to reach the 5\% error bound relative to full-data processing, averaged over all target categories within each dataset. For the \textit{Movie} dataset, the \textit{Adjust} method achieves this threshold using only 13.75\% of the full time on average across classes, substantially outperforming \textit{No Adjust} (23.71\%) and \textit{Random} (21.80\%). 
Similarly, in the \textit{BBC News} experiment, the average time proportions are 10.82\%, 14.58\%, and 39.40\% for \textit{Adjust}, \textit{No Adjust}, and \textit{Random}, respectively. 
\vspace{-0.2in}

\subsection{Impact of Hyper-parameters}
\label{subsec:val-hyper}

In this section, we investigate the impact of key hyper-parameters on the performance of \toolname.
We aim to provide empirical guidance for choosing appropriate hyper-parameter settings to maximize estimation accuracy.

\vspace{-0.2in}

\begin{figure}[ht]
    \centering
    \includegraphics[width=0.6\linewidth]{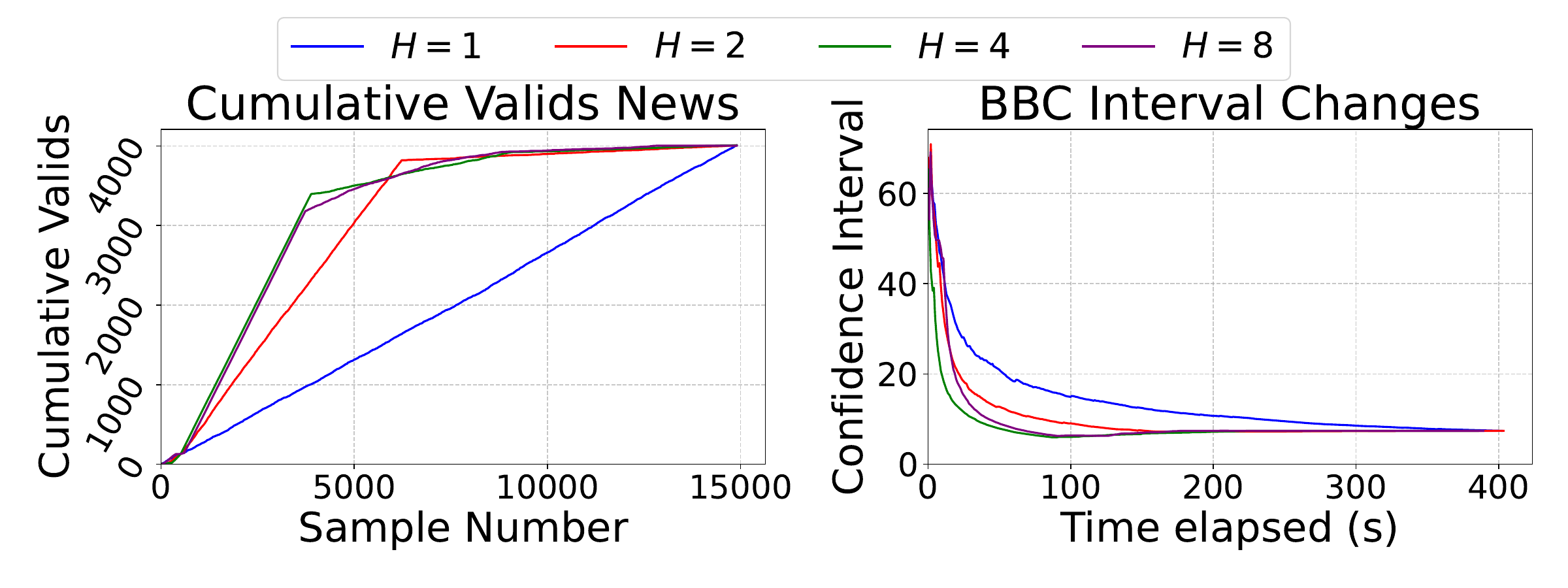}
    \vspace{-10pt}
    \caption{Impact of the number of strata on convergence of Confidence Intervals in filtering scenario. The query is to filter items belong to the \textit{sports} category in \textit{BBC News}.}
    \label{fig:exp_hyper_para_filter}
\vspace{-0.3in}
\end{figure}

For the filtering scenario, we investigate the impact of the number of strata on the convergence of confidence intervals. The results, shown in Fig.~\ref{fig:exp_hyper_para_filter}, demonstrate that even with as few as two strata, semantic sampling already achieves a significant acceleration in convergence. 
This improvement can be attributed to the strong semantic separability of the \textit{BBC News} dataset, which enables the clustering algorithm to effectively distinguish among different semantic groups. Furthermore, as the number of strata increases to four or five, reflecting the five major news categories present in the dataset, convergence further improves, yielding more accurate estimates with reduced sampling effort.
\vspace{-0.2in}

\begin{figure*}[ht]
    \centering
    \includegraphics[width=1.0\linewidth]{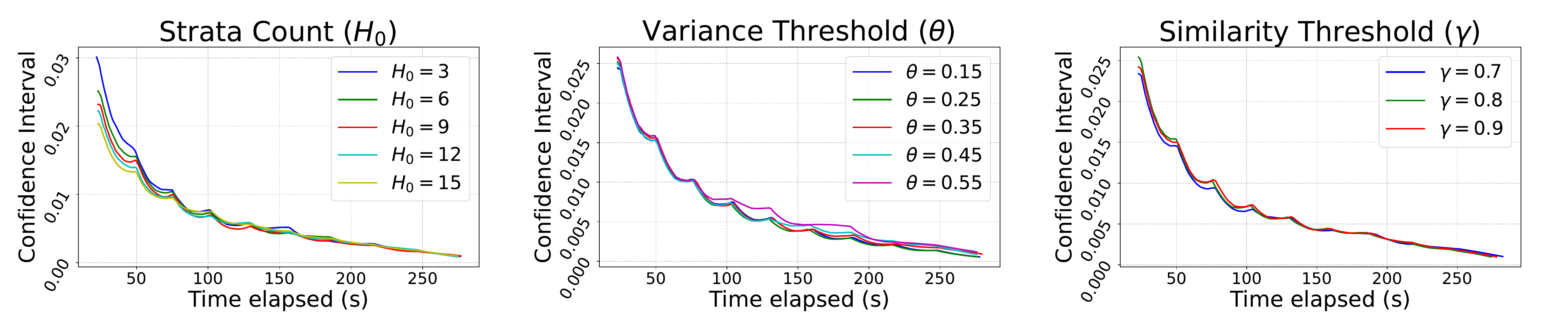}
    \vspace{-15pt}
    \caption{Impact of hyper-parameter settings on convergence in group-by scenario. The query is designed to compute the ratio of different categories (Positive, Negative and Neutral) within the \textit{Movie} dataset.}
    \label{fig:exp_hyper_para_groupby}
\vspace{-0.2in}
\end{figure*}

For the group aggregation scenario, we further investigate the effects of key hyper-parameters on convergence behavior. This experiment is based on the previously described \textit{Movie} dataset, where the objective is to compute the proportion for each semantic group. The results, presented in Fig.~\ref{fig:exp_hyper_para_groupby}, show that employing a larger number of initial strata typically contributes to faster convergence in the early sampling phases. This observation can be explained by the enhanced semantic resolution that a greater number of strata provides, allowing for more accurate group-wise estimates. The second plot in Fig.~\ref{fig:exp_hyper_para_groupby} illustrates the impact of the variance threshold on convergence: when this threshold is set too high (e.g., 0.55), it may prohibit finer-grained strata formation and consequently slow down convergence. In contrast, the third plot reveals that the similarity threshold exerts a relatively small influence on the overall convergence performance.
\vspace{-0.2in}


\subsection{Scalability}
\vspace{-0.1in}

\label{subsec:scalability}

\begin{figure}[ht]
    \centering
    \includegraphics[width=0.7\linewidth]{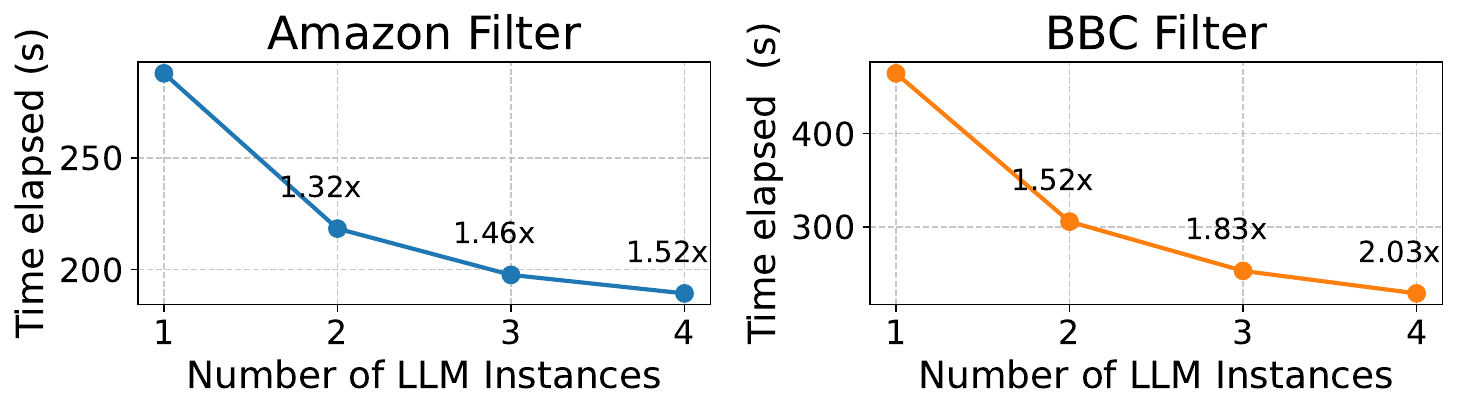}
    \vspace{-5pt}
    \caption{System scalability evaluation.}
    \label{fig:exp_scalability}
\vspace{-0.25in}
\end{figure}

We evaluate the scalability of our system by increasing the number of LLM serving instances to assess how performance improves. 
We deploy our system using LiteLLM~\cite{litellm} as the serving layer, scaling from 1 to 4 instances with identical hardware specifications, and measure the execution time required to reach the same error bound in two representative workloads: filtering and aggregation on \textit{Amazon Product} and \textit{BBC News}.
Fig.~\ref{fig:exp_scalability} illustrates the execution time with respect to the number of LLM serving instances. 
As shown in the figure, both workloads benefit from scaling, but the incremental gains diminish as additional instances are added. 
With 2, 3, and 4 serving instances, we observe speedups of 1.32$\times$, 1.46$\times$, and 1.52$\times$ for the \textit{Amazon Product} workload, and a more scalable 1.52$\times$, 1.83$\times$, and 2.03$\times$ for the \textit{BBC News} workload. These results confirm that our architecture scales effectively. The sub-linear trend arises mainly from LiteLLM’s default routing and batching configuration, which is not yet tuned for our workload. With optimized settings, scalability could approach the linear ideal.


\vspace{-0.2in}

\vspace{-0.2in}

\section{Conclusion}

In this paper, we present {\toolname}, a novel framework that integrates large language models with online aggregation to enable scalable, interactive analytics over unstructured data. By transforming unstructured inputs into structured representations through LLM-based extraction, embedding them for semantic indexing, and applying approximate query processing via online aggregation, {\toolname} addresses long-standing challenges in latency, scalability, and semantic ambiguity. 
Experimental results across diverse domains demonstrate that {\toolname} enables faster time-to-insight and higher-quality exploration compared to existing approaches. 


\begin{credits}
\subsubsection{\ackname} This work was supported by the Fundamental Research Funds for the Central Universities and the Research Funds of Renmin University of China (24XNKJ22), the Beijing Natural Science Foundation (L247027), National Natural Science Foundation of China (No.62272466, No.U2436209, and No.U24A20233). The computing resource was supported by PCC of RUC.
\subsubsection{\discintname}

\end{credits}
%
%
%
\bibliographystyle{splncs04}
\bibliography{reference}

\end{document}